\documentclass{article}
\usepackage[hyphens]{url}
\usepackage[utf8]{inputenc}
\usepackage{graphicx}
\usepackage{authblk}
\usepackage[english]{babel}
\usepackage{setspace}
\usepackage{fancyhdr}
\usepackage{array}
\usepackage[margin=1in]{geometry}
\usepackage{float}
\usepackage{amsmath}
\usepackage{amssymb}
\usepackage{bbm}
\usepackage{bm}
\usepackage{scalerel,stackengine}
\usepackage[final]{pdfpages}
\usepackage{longtable}
\usepackage{chngcntr}
\usepackage{placeins}
\usepackage{microtype}
\usepackage{tikz}
\usepackage{makecell}
\usepackage{hyperref}
\usepackage{subcaption}
\usepackage{rotating}
\usepackage{soul}
\usepackage{tabularx} 
\usepackage{pdfpages}
\usepackage[square,numbers]{natbib}
\usepackage{multirow}
\usepackage{listings} 
\usepackage{multirow}
\usepackage{enumitem}  
\usepackage{graphicx}
\usepackage{subcaption} 
\usepackage{lscape} 

\hypersetup{
    colorlinks = true,
    citecolor = {blue},
    urlcolor = {blue},
    menucolor = {blue},
    linkcolor = {blue}
}

\usetikzlibrary{shapes, decorations, positioning, calc}
\makeatletter
\patchcmd{\@maketitle}{\LARGE \@title}{\fontsize{18}{20}\selectfont\textbf{\@title}}{}{}
\makeatother

\title{Treatment-control comparisons in platform trials including non-concurrent controls}  

\author[1]{Marta Bofill Roig\thanks{marta.bofillroig@meduniwien.ac.at}} 
\author[1]{Pavla Krotka}
\author[2]{Katharina Hees}
\author[1]{Franz Koenig}
\author[3]{Dominic Magirr}
\author[4]{Peter Jacko}
\author[4]{Tom Parke}
\author[1]{Martin Posch}
\author[]{on behalf of EU-PEARL (EU Patient-cEntric clinicAl tRial pLatforms) Consortium} 

\affil[1]{Center for Medical Data Science, Medical University of Vienna, Vienna, Austria}
\affil[2]{Section of Data Science and Methods, Paul-Ehrlich-Institut, Germany}
\affil[3]{Advanced Methodology and Data Science, Novartis Pharma AG, Switzerland}
\affil[4]{Berry Consultants, UK}
\affil[5]{Lancaster University, UK} 

\date{}         
\begin{document}

\maketitle

\begin{abstract}
	{Shared controls in platform trials comprise concurrent and non-concurrent controls. For a given experimental arm, non-concurrent controls refer to data from patients allocated to the control arm before the arm enters the trial. The use of non-concurrent controls in the analysis is attractive because it may increase the trial’s power of testing treatment differences while decreasing the sample size. However, since arms are added sequentially in the trial, randomization occurs at different times, which can introduce bias in the estimates due to time trends.
		In this article, we present methods to incorporate non-concurrent control data in treatment-control comparisons allowing for time trends. We focus mainly on frequentist approaches that model the time trend and Bayesian strategies that limit the borrowing level depending on the heterogeneity between concurrent and non-concurrent controls. We examine the impact of time trends, overlap between experimental treatment arms and entry times of arms in the trial on the operating characteristics of treatment effect estimators for each method under different patterns for the time trends. We argue under which conditions the methods lead to type 1 error control and discuss the gain in power compared to trials only using concurrent controls by means of a simulation study in which methods are compared.}
	{Platform trials; External controls; Time trends}
\end{abstract}


\section{Introduction}

Platform trials offer a highly efficient way to evaluate multiple treatments using a single infrastructure \cite{Woodcock2017,Berry2015,meyer2020evolution, koenig2024current}. By allowing different experimental treatments to enter and exit at different times, platform trials facilitate a faster evaluation of new treatments as the efficacy of new treatments can be investigated as soon as they become available \cite{Saville2016}. This approach optimizes resource utilization and enables a more rapid and comprehensive assessment of potential treatments \cite{Collignon2021,Collignon2020}.
Platform trials usually consider a shared control on the basis of which to assess the effectiveness of treatments. For treatment arms entering later, control data are distinguished between concurrent data, referring to data from patients in the control arm randomised concurrently in time with the entering arm, and non-concurrent data, referring to control data from patients in the trial randomised before the arm entered. The use of non-concurrent data in the analysis comparing the efficacy of treatments against control has been widely discussed in recent years \cite{bofill2023scoping,Dodd2021,Sridhara2021a,Park2022,Stallard2020, BofillRoig2022,Marschner2022,FDA2023masterprotocol}.

Several approaches have been recently suggested to incorporate non-concurrent controls in platform trials \cite{bofill2023scoping}.  Frequentist and Bayesian modelling approaches that utilize non-concurrent control data and adjust for time trends have been proposed. In trials with binary and continuous endpoints, frequentist methods have been proposed with  adding time as a covariate to the regression model to adjust for temporal changes \cite{Lee2020,BofillKrotka2022}. These adjustments were presented in the context of a platform trial with two experimental arms and a shared control. 
For trials with binary endpoints, Bayesian strategies include the Time Machine approach, which considers a Bayesian generalized linear model that smooths the control response over time \cite{Saville2022}. 
In the context of historical controls, the meta-analytic-predictive (MAP) Prior approach was proposed as a Bayesian borrowing method, that models the between-trial variation \cite{Schmidli2014,Weber2021}. The MAP approach performs a prediction of the control effect in the trial from historical control data using random-effects meta-analytic methods.

Simulation studies are an important tool to assess the operating characteristics of trials using external controls \cite{KoppSchneider2023}  and also recommended by regulatory authorities \cite{EMA2022QA}. Although different comparative studies have been carried out through simulations to propose and/or evaluate the characteristics of different methods using external or historical controls \cite{Gotte2023, Kopp-Schneider2020, Hupf2021,Qi2022,Jiao2019,Ventz2021,Viele2014}, there are so far no simulation studies comparing the recently published methods  for the use of non-concurrent controls in platform trials. 

In this work, we consider trials with continuous data. We aim to extend existing methods for treatment-control comparisons incorporating non-concurrent control data and compare them in a simulation study. 
We focus mainly on frequentist and Bayesian modelling approaches that model the time trend as well as on Bayesian strategies that limit the borrowing level depending on the heterogeneity between concurrent and non-concurrent controls.   More precisely, we consider frequentist model-based adjustments proposed in \cite{BofillKrotka2022} and extend them for a more general setting of platform trials with $K$ experimental arms ($K\geq 2$), adapt Bayesian meta-analytic-predictive approaches commonly used in the context of incorporating historical data into the analysis, and extend the Bayesian Time Machine proposed in \cite{Saville2022} for platform trials with binary endpoints to continuous endpoints.  
We evaluate the statistical power and the type 1 error rate of the methods for individual treatment-control comparisons and compare them through an extensive simulation study. Specifically, we assess the statistical properties of using non-concurrent controls when using these methods over a wide range of settings, including different time trend patterns and scenarios with equal or different strengths of the time trend across arms, as well as varying the number of experimental arms added later, the frequency of new arms entering and the overlapping period between experimental arms throughout the trial.  

The paper is organised as follows. In Section \ref{sect_methods}, we describe the trial design and main notation, and present several methods for comparing treatment against control in platform trials using non-concurrent controls. In Section \ref{sect_simstudy}, we compare the different methods in a simulation study.  
We finish with conclusions and a discussion in Section \ref{sect_discuss}.  

\section{Methods to use non-concurrent controls}\label{sect_methods}

Consider a randomized controlled trial evaluating the efficacy of multiple experimental treatments compared to a control treatment. 
Suppose that the experimental treatment arms enter sequentially into the trial up to a total number of $K$ different experimental treatment arms. Participants are randomized equally between control and active treatment arms upon recruitment, where active refers to open for enrollment.  
Suppose that the sample sizes of experimental treatment arms are equal to $n$ and the total sample size in the platform trial equals $N$. 
We denote by $j \in \mathcal{J}=\{1,\cdots,N\}$ the participant index and by $k \in \mathcal{K}=\{0,\cdots,K\}$ the treatment indicator,  $k=0$ indicating the control treatment and $k>0$ indicating the experimental treatments ordered by entry times.

Denote by $y_{j}$ the response for the $j$-th participant, assumed to be continuous. For simplicity, we assume that the response is obtained at the same time as the participant enters the trial. We denote the treatment effect size for treatment $k$ compared to the control by $\theta_k$ ($k\in \mathcal{K}$) and consider the respective hypothesis testing problem:
\begin{eqnarray*}
	H_{0,k}: && \theta_k \leq 0\\
	H_{1,k}: && \theta_k>0
\end{eqnarray*}


In this work, we focus on the inference on those arms that enter the trial when it is already running, and therefore, for those where non-concurrent control data is available.  
We aim to compare experimental treatments against the control as soon as this experimental treatment arm leaves the trial. 
In what follows, we present several approaches to test $H_{0,k}$ by incorporating the non-concurrent control data in the treatment-control analysis. In particular, we consider frequentist model-based adjustments in Section \ref{sect_freqmods}, the Bayesian Time Machine in Section \ref{sect_tmachine} and the Meta-Analytic-Predictive (MAP) approach in Section \ref{sect_map}. 
The methods differ with respect to the data used for the analyses and in whether the time component of the trial is considered to adjust for potential trends and, if yes, how it is used. For readability, the notation and concepts specific to each method are introduced in its corresponding section.


\subsection{Frequentist model-based adjustments} \label{sect_freqmods}

To evaluate the efficacy of experimental treatment $k$ compared to the control,  we consider a regression model, where time is modelled as a step function as proposed in  \cite{BofillKrotka2022}. To incorporate time into the model, we divide the platform trial duration into intervals where there is no change in the number of active arms. A new time period starts when an experimental treatment arm is added (starts being active) or removed (stops being active) from the platform. 
Assume that the study has $S$ periods ($S>1$), and denote by $s \in \mathcal{S}= \{ 1, \cdots, S \}$ the period indicator.  The observed data is $\{(y_j, k_j, s_j)$, $j=1, \cdots, N\}$, where $k_j$ is the treatment of participant $j$; and $s_j$ represents the period at which the participant  $j$ enters. For a given treatment $k\in\mathcal{K}$, we denote by $S_k$ the period in which arm $k$ finished, and by $\mathcal{K}_{S_k} \subseteq \{1, ..., K\}$ the set of experimental treatments in the platform prior or up to $S_k$. 

\begin{figure}[h!] 
	\centering
	\includegraphics[width=0.45\textwidth]{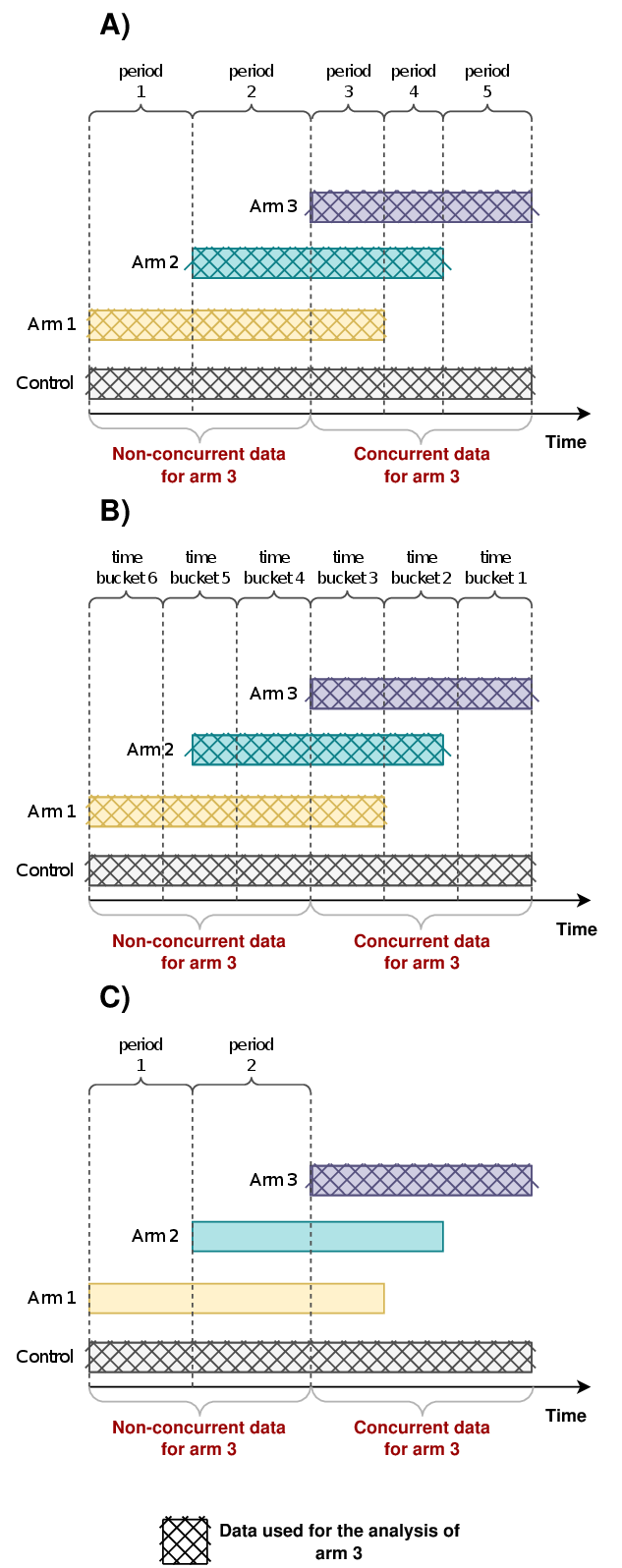} 
	\caption{Platform trial with three experimental treatment arms entering sequentially over time. The figure marks concurrent and non-concurrent control data for arm 3 and indicates with crossbars the data used to analyse the efficacy of treatment 3 versus control when using A) the frequentist regression model, B) the Bayesian Time Machine, and C) the MAP Prior approach. }
	\label{fig:data_combined}
\end{figure}

Figure \ref{fig:data_combined}-A) shows a platform trial with three experimental treatment arms and a control, where treatment arms enter sequentially, resulting in a platform trial with five periods. If the treatment under current evaluation is the treatment administrated in Arm 3, the non-concurrent data corresponds to the data from periods 1 and 2.  

To test $H_{0k}$, we consider 
the t-test coming from a frequentist linear model which estimates the treatment effects of experimental arms in the trial up to period $S_k$
and includes time  as a factor in the analysis of the trial. The model is given by: 
\begin{equation} \label{eq_act}
	E(y_j) = \eta_0  + \sum_{k' \in \mathcal{K}_{S_k}} \theta_{k'} \cdot I(k_j=k') + \sum_{s=2}^{S_k} \tau_s \cdot I(s_j=s)
\end{equation} 
where $\eta_0$ denotes the response in the control group in the first period, $\theta_k$ represents the effect of the treatment $k$ compared to control, and $\tau_s$ denotes a step-wise time effect between periods 1 and $s$. 

This approach models time using a step function, and implicitly assumes: (i) the period-time effect is the same for all arms in the platform trial, (ii) the time effect is constant in each period and (iii) this effect is additive in the model scale. 
As we test the efficacy of arm $k$ when this leaves the platform, we fit the model using all data from the trial until experimental treatment arm $k$ leaves the platform, i.e., $\{ (y_j, k_j, s_j), j=1,...,N : s_j \le S_k \}$. Hence, data from all experimental arms that were active at some point up to  $S_k$ contribute to estimating the effect of time by means of $\tau_s$ ($s=2, ..., S_k$).  Figure \ref{fig:data_combined}-A) illustrates the data used to fit the model and thus to adjust for time trends.


\subsection{Bayesian Time Machine modelling} \label{sect_tmachine}

Saville et al.  \cite{Saville2022} proposed time-adjusted analyses to model potential temporal drifts over the trial. The model, so-called Bayesian Time Machine, is built on the basis of a generalized linear model to evaluate multiple experimental treatment arms versus a control in trials with binary endpoints, extended here for continuous endpoints.

In the Time Machine model, the time is incorporated differently than in the frequentist model. Instead of considering the variable ``period'', the time is adjusted using equal-sized time intervals, called ``time buckets'', indexed backwards from the most recent time interval to the beginning of the
trial. Figure \ref{fig:data_combined}-B) illustrates a platform trial with three experimental treatment arms and a control, splitting the time duration into time buckets. 

Aiming at comparing a given treatment $k\in\mathcal{K}$ against control and assuming that the trial has a total of $C_k$ time buckets ($C_k>1$) when arm $k$ leaves the trial, we denote by $c \in \mathcal{C}= \{ 1, \cdots, C_k \}$ the bucket indicator, where $c=1$ corresponds to the last time bucket in which treatment $k$ is active in the trial and $c=C_k$ denotes the beginning of the trial.  
Analogously to the previous section, the observed data is $\{(y_j, k_j, c_j)$, $j=1, \cdots, N\}$, where $y_j$ and $k_j$ are the continuous response and treatment indicator for participant $j$ as before, but now the time information is given by $c_j$ that represents the time bucket at which the participant  $j$ enters. 
Denote by $\mathcal{K}_{C_k} \subseteq \{1, ..., K\}$ the set of active treatments in the platform prior to or up to $C_k$.

We extend the model to trials with continuous endpoints as follows:
\begin{eqnarray} 
	Y_j &=& E(Y_j) + \epsilon_j \\
	E(Y_j) &=&  \eta_0 + \sum_{k' \in \mathcal{K}_{C_k}} \theta_{k'} \cdot I(k_j=k') + \sum_{c=2}^{C_k}  \omega_{c} \cdot I(c_j=c) 
\end{eqnarray}
where $\eta_0$ is the intercept and $\theta_{k}$ are the treatment effects, with typically (nearly) non-informative prior distributions that depend on the scale of the data:
\begin{eqnarray*}
	\eta_0 &\sim& N(0,\sigma^2_{\eta_0})  \\
	\theta_k &\sim& N(0,\sigma^2_\theta)   
\end{eqnarray*}  
The parameter $\omega_{c}$ is the increment predictor for the time bucket and quantifies the drift over time, where $\omega_{1}$ corresponds to the most recent time interval. Similarly, as in \cite{Saville2022}, for every previous time interval, the time parameter is modeled with the following Bayesian second-order normal dynamic linear model: 
\begin{eqnarray}\label{NDLM}
	\omega_1 &=& 0 \\
	\omega_{2} &\sim& N(0, 1/\tau)  \\
	\omega_{c} &\sim& N(2\omega_{c-1} - \omega_{c-2}, 1/\tau) , \ 3\le c\le C_k   
\end{eqnarray}
The precision parameter $\tau$ is the inverse of the variance and specifies the degree of smoothing over time intervals. A hyperprior distribution is then specified as follows:
\begin{eqnarray}
	\label{tau}
	\tau &\sim& \text{Gamma}(a_{\tau}, b_{\tau})  
\end{eqnarray} 
The precision of the individual participant responses, $\epsilon_j$, is also assumed to have a Gamma hyperprior distribution $\epsilon_j \sim N (0, 1 / \tau_Y)$ with $\tau_Y \sim \text{Gamma}(a_Y, b_Y)$.

Analogously as for the frequentist model, and as illustrated in Figure \ref{fig:data_combined}-B), we fit the Time Machine model using all data from the trial until the experimental treatment arm $k$ leaves the platform, i.e., $\{(y_j, k_j, c_j)$, $j=1, \cdots, N : c_j \leq C_k \}$.

A non-informative prior may be used for $\tau_Y$. 
However, using this type of prior also for $\tau$ is generally not appropriate when the number of time intervals may be small \cite{Gelman2006}. Rather, a weakly informative approach is recommended, whereby the bulk of the prior density covers an a-priori plausible region of the parameter space.  
To calibrate such a weakly informative prior, it can be helpful to think about plausible changes in $E(Y)$ between time buckets $c$ and $c-1$. 
All such changes $\omega_{c-1} - \omega_{c}$ have a standard deviation of $\tau^{-1/2}$. The first change $\omega_{C-1} - \omega_{C}$ has a mean of zero, while all subsequent changes have a mean equal to the previous change. We could then consider what is our most plausible value for $\tau^{-1/2}$, denoted $D_{\text{Expected}}$, and what do we consider a very large value of $\tau^{-1/2}$, denoted $D_{\text{Maximum}}$, such that we have only a small belief,  $ \iota$, that $\tau^{-1/2} > D_{\text{Maximum}}$. That is, we solve
\begin{eqnarray}\label{d_exp}
	E(\tau) &=& 1/D_{\text{Expected}}^2 \\ \label{d_max}
	P(\tau &<&  1/D_{\text{Maximum}}^2) = \iota
\end{eqnarray} 
for $a_{\tau}$ and $b_{\tau}$, given some small value of $\iota$, e.g. 0.01.

As in the frequentist model, the Time Machine models time using a step function and assumes that the period-time effect is the same for all arms in the platform trial and that this effect is additive in the model scale. Another similarity between the two approaches is the fact that both models use all available data to estimate the effect of time.  

\subsection{Meta-Analytic Predictive prior approaches}\label{sect_map}

The Meta-Analytic Predictive (MAP) prior approach was proposed as a method to involve data from multiple historical studies \cite{Schmidli2014} in the final analysis of a current clinical trial. It aims at summarising relevant sources of information (data from historical controls) while accounting for between-trial heterogeneity. The resulting distribution, the MAP prior, is then used as an informative prior for the concurrent controls in the final analysis. 
The MAP prior is derived using a random-effects meta-analysis model with a subsequent prediction for the control mean of a future trial.

In this work, we consider the MAP prior as an approach to perform the treatment-control comparisons in platform trials using non-concurrent controls. More precicesly, in the context of platform trials, the MAP approach can be used to derive the prior distribution for the control response in the concurrent periods by combining the control information from the non-concurrent periods with an initial non-informative prior. 
To evaluate the efficacy of treatment $k$ compared to the control, consider the non-concurrent control responses
$Y_{NCC} = \{ y_j : k_j=0, s_j\leq  \Bar{S}_k\}$ 
where $\Bar{S}_k$ denotes the period preceding $k$ entering the trial. 
Figure \ref{fig:data_combined}-C) illustrates the periods (defined in Sect. \ref{sect_freqmods}) preceding Arm $3$.  
Note that $Y_{NCC}$ are the non-concurrent controls with respect to experimental treatment arm $k$, but we omit the dependence on $k$ in the notation for simplicity.

For $y_j\in Y_{NCC}$, let
$y_j|\eta_{s_j} \sim f(\eta_{s_j})$, where $f$ is the likelihood function. 
In order to borrow strength from the source information from the different periods, we consider a hierarchical model for the control response in period $s$ that links the parameters from the different  periods:
\begin{eqnarray}
	\eta_s &=& \beta + \nu_s \\
	\nu_s &\sim& N(0, \tau^2)
\end{eqnarray} 
where $s=1, ..., \Bar{S}_k$, omitting here the subindex $j$ in $s_j$ for simplicity. In this model,  $\beta$ is the population mean in the control, and $\nu_s$ is the variability introduced by the periods with mean 0 and standard deviation $\tau$, which can be interpreted as the between-period heterogeneity. For $\beta$ and $\tau$, the following hyperprior distributions are assumed: 
\begin{eqnarray}
	\label{MAP_sigmabeta}
	\beta &\sim& N(0, \sigma^2_{\beta}) \\
	\label{MAP_sigmatau}
	\tau &\sim& \text{HalfNormal}(0, \sigma^2_{\tau})
\end{eqnarray} 
The data from period $\Bar{S}_k+1$ and beyond is not available before experimental treatment arm $k$ is included in the platform trial. Therefore, the posterior of the parameters is based on non-concurrent data only. The prior information on the concurrent controls is the posterior for the above-specified model, called the MAP prior, that is
$$
p_{\text{MAP}}(\eta_{S_k}) = P(\eta_{S_k}|Y_{NCC})
$$
Once the concurrent control data for treatment $k$ is available, the posterior for $\eta_{S_k}$  can be obtained as $ p(\eta_{S_k}|Y_{CC}) \propto p(Y_{CC}|\eta_{S_k}) \cdot p_{\text{MAP}}(\eta_{S_k})$, where $Y_{CC}$ are the concurrent controls, that is, $Y_{CC} = \{ y_j : k_j=0, \Bar{S}_k< s_j\leq  S_k\}$.

The MAP prior can be robustified to avoid prior-data conflicts. This is achieved by adding a weakly-informative mixture component $p_{\text{non-inf}}$, resulting in the prior distribution
\begin{equation}
	p_{\text{rMAP}}=(1-a_R)p_{\text{MAP}}+a_R p_{\text{non-inf}}
\end{equation}
where $a_R$ is a weight that can be interpreted as the degree of scepticism towards borrowing strength.

Note that the MAP approach is conceptually different from the previous two models. Here, the hierarchical model (and thus the use of non-concurrent controls) is only to build the prior of the concurrent controls. In addition, data from other experimental treatment arms are not used in this case (see Figure \ref{fig:data_combined}-C)) for a representation of the data usage).



\section{Simulation study} \label{sect_simstudy}

We simulated platform trials evaluating the efficacy of $K$ experimental treatment arms compared to a shared control. Arm $k$ ($k>1$) enters after $d_k$ participants have been recruited to the trial and $d_1=0$.   
To investigate the properties of the methods for utilising non-concurrent controls under different situations, we considered three trial settings explained in Sect. \ref{sec:simsett}.  The settings differ from each other in the number of experimental arms, $K$, and the overlaps between arms, $\mathbf{d} = (d_1,...,d_K)$. 
In all three settings, we assume all experimental arms were equal sized with sample sizes of 250, equal allocation among control and treatment in each period as well as block randomisation per period. 
Note that the sample size for the control arm (and thus for the overall trial) varies depending on the entry pattern and overlapping between experimental treatment arms. 

We compare the performance of the frequentist regression model (Section \ref{sect_freqmods}),  Bayesian Time Machine (Sect. \ref{sect_tmachine}) and MAP prior (Sect. \ref{sect_map}) in terms of individual power and type 1 error rate. For comparative purposes, we also considered  the so-called separate analysis   (t-test  using concurrent controls), and the so-called pooled analysis (t-test pooling concurrent and non-concurrent control data without adjustments). We describe the choice of the priors for the Time Machine and MAP approaches in Section \ref{sec:simmeth}.

We performed all computations using \texttt{R} software. 
For the simulation and analysis, we used the  \texttt{NCC} R package \cite{krotka2023}. The code to reproduce the results is available at  

\url{https://github.com/pavlakrotka/NCC_MethodsComp}.

\subsection{Data generation and trial settings}\label{sec:simsett}

We simulated trials with continuous data using the generating model 
\begin{equation} \label{model_datag} 
	E(Y_j) = \eta_0 + \sum_{k=1,...,K} \theta_k \cdot I(k_j = k) + f(t_{j}),
\end{equation}
where  $Y_j$, $\eta_0$ and $\theta_{k_j}$ refer to the continuous response, the control response and treatment effects, respectively. We furthermore assume that the error terms in the responses are identically and independently normally distributed with zero mean and homoscedastic variances equal to 1.  

We assumed effect sizes of $\theta_{k} = 0.25, k=1,...,K$ for the treatment-control comparisons under the alternative hypothesis, and a response of 0 in the control. When evaluating under the null hypothesis, we simulate all experimental treatment arms under the null hypothesis, while when evaluating under the alternative hypothesis, we assume all experimental treatment arms under the alternative hypothesis.
The chosen sample and effect sizes lead to 80\% power for the treatment-control comparison using a separate analysis (one-sided t-test at 2.5\% significance level using only concurrent controls). 

The term $f(\cdot)$ represents the time trend function and $t_j$ is the calendar time when participant $j$ is enrolled in the trial. We assume that one and only one patient enters the trial at a particular calendar time, so that $t_j=j$ for all $j$.
Similarly as in \cite{BofillKrotka2022}, we considered three time trends patterns: 
\begin{itemize}
	\item Stepwise time trend: $f(j) = \lambda_{k_j} \cdot (c_j - 1)$, where $c_j$ is the number of  experimental treatment arms have already entered the ongoing trial when participant $j$ was enrolled
	\item Linear time trend: $f(j) = \lambda_{k_j} \cdot \frac{j-1}{N-1}$, where $N$ is the total sample size in the trial
	\item Inverted-U time trend: $f(j) = \lambda \cdot \frac{j-1}{N-1}$ for  $j \leq N_p$, and $f(j) = -\lambda \cdot \frac{j-N_p}{N-1} + \lambda \cdot \frac{N_p-1}{N-1}$ for $j > N_p$, where $N_p$ denotes the sample size at which the form of the time trends changes. $N_p$ is set to approximately $N/2$ so that the peak is always approximately in the middle of the trial.
\end{itemize}
where the parameter $ \lambda_{k_j}$ quantifies the strength of the time trend. Note that the functional form of the time trend is assumed to be equal across arms. If the strength of the time trend is equal across arms, then we will say that it satisfies the assumption of equal time trends. 
Figure \ref{fig:trend_patterns} illustrates the time trend patterns. Note that the stepwise time trend is more severe in terms of the change in mean responses over time. Also, note that the linear time trend is more pronounced for a specific arm when there are more arms open compared to when there are fewer arms open, given that in the first case, the arm under question runs for a longer time than in the second case. 

\begin{figure}[h!]
	\centering
	\includegraphics[width=0.5\textwidth]{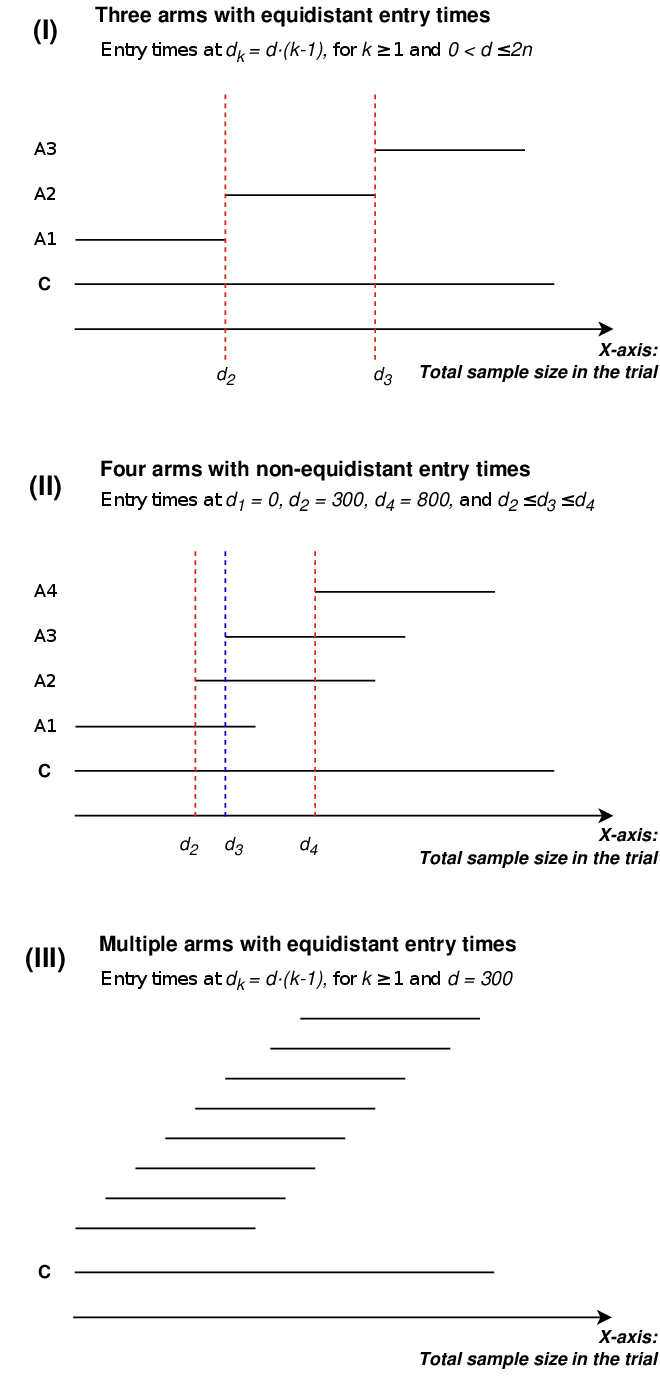}
	\caption{Graphical visualisation of platform trials used for three settings being investigated by clinical trial simulations: (I) trial with three experimental arms and equidistant
		entry times, varying the difference in entry times between consecutive arms ($d$); (II) trial with four arms and non-equidistant entry times, varying the entry time of arm 3 ($d_3$); (III) trial with ten arms and equidistant entry times varying the strength of the time trends for the control and arm 10, assumed to be equal ($\lambda_0=\lambda_{10})$.}
	\label{fig:simdesign}
\end{figure}

\begin{figure}[h!]
	\centering
	\includegraphics[width=0.5\textwidth]{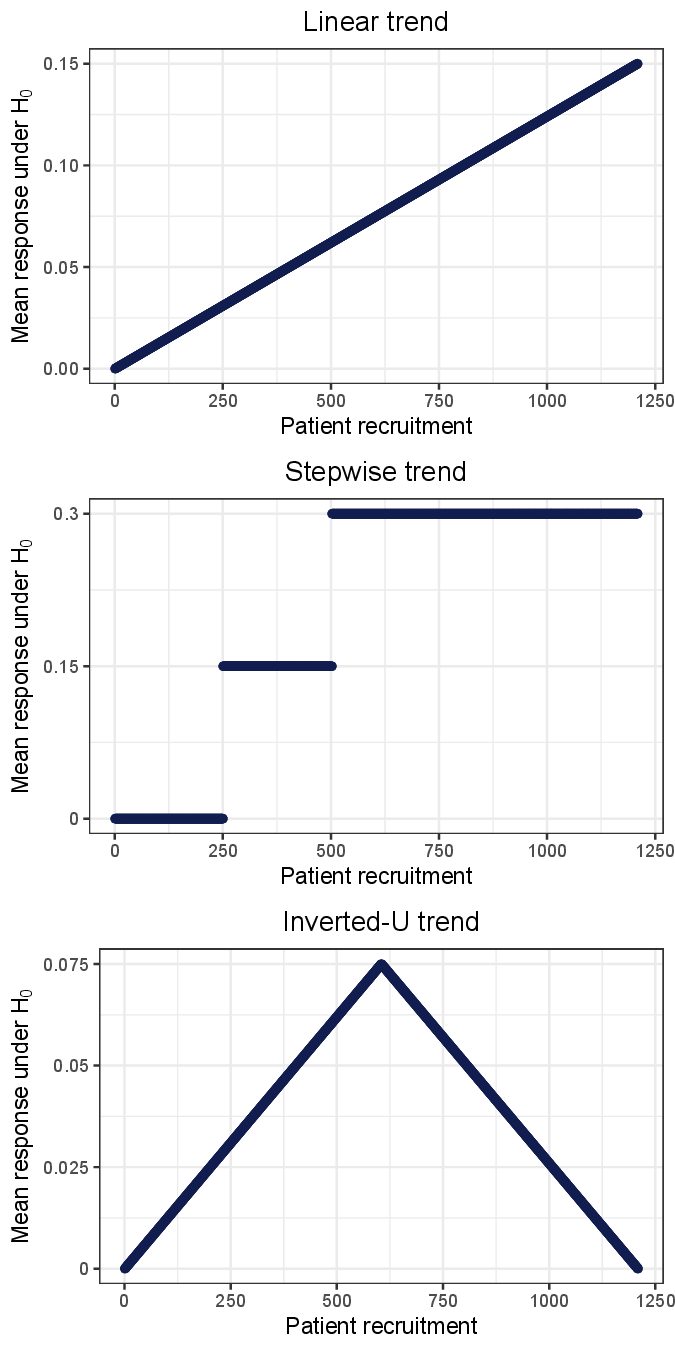}
	\caption{Mean responses under the null hypothesis under time trends of different patterns and strength of $\lambda_k = 0.15$, $\forall k$. }
	\label{fig:trend_patterns}
\end{figure}

To investigate the impact of time trends, entry times, and overlaps between arms, we consider the following platform trial settings, varying, in each of them, different elements to understand their implications in the analysis: 
\begin{enumerate} 
	\item[] \textbf{Setting I: Platform trial with three experimental arms and equidistant entry times}. We explore the effect of the overlap between arms in trials with $K=3$ experimental arms, in settings with equal time trends and different time trends.
	
	We examine a platform trial with three experimental treatment arms, where treatment arm $k$ enters after every $d_k = d \cdot (k-1)$ participants have joined the trial. We consider five options for $d$: $d = 0, \ 125,  \ 250,  \ 375$ and $500$, resulting in platform trials with different overlaps between arms.      
	We consider linear, stepwise and inverted-U time trends. As for the strength of the time trend, we consider time trends that are equal across all arms ($\lambda_k=\lambda_0$, $\forall k\geq1$), that differ from the control in arm 1 ($\lambda_3=\lambda_2=\lambda_0 \neq \lambda_1$), and that differ  from the control in arms 1 and 2 ($\lambda_3=\lambda_0 \neq \lambda_1=\lambda_2$). We investigate the type 1 error and power of comparing arm 3 against control in cases with different time trends. Note that when $d=0$, all arms enter and finish at the same time, so we are in a classical multi-arm trial setting, while if $d=500$, a new treatment arm enters when the preceding one ends.

	\item[] \textbf{Setting II: Platform trial with four arms and non-equidistant entry times}. We investigate the impact of different entry times in trials with $K=4$ arms and equal stepwise time trends. 
	
	Focusing on evaluating the treatment efficacy of arm 4 against control, we examine a platform trial with four experimental treatment arms, where treatment arms 2 and 4 enter after every $d_2=300$ and $d_4=800$ participants have been recruited, respectively. For the timing of adding the treatment arm 3, we consider $d_3 = 300, \ 425, \ 550, \ 675, \ 800$.
	
	\item[] \textbf{Setting III: Platform trials with multiple arms and equidistant entry times} We explore the operating characteristics of platform trials with $K=10$ arms.     
	We investigate the impact of different time trends between arms with potentially random strength. 
	
	We consider a platform trial with $K=10$ experimental treatment arms, where treatment arm $k$ enters after every $300 \cdot (k-1)$ participants have been recruited to the trial. 
	We use linear time trends only and focus on evaluating the efficacy of arm 10 against control. Here, the time trend in arm 10 is equal to the control group ($\lambda_0 = \lambda_{10}$), for which we consider different values. The time trend in the remaining treatment arms is sampled from $\lambda_k \sim N(\lambda_0, 0.5), \forall k \in \{1,\ldots,9\}$.    
\end{enumerate}
We illustrate the settings in Figure  \ref{fig:simdesign} and summarise the investigated aspect and considered parameters in Table \ref{table_settings}.
For each considered scenario, we used 10,000 simulation replicates.

\begin{landscape}
	\begin{table}[]
		\resizebox{22cm}{!}{\begin{tabular}{|c|c|c|c|c|c|c|}
				\hline
				\textbf{Setting} & \textbf{Investigated aspect} & $K$ & $d$ & $\lambda$ & \begin{tabular}[c]{@{}c@{}}\textbf{Trend}\\ \textbf{pattern}\end{tabular} & \begin{tabular}[c]{@{}c@{}}\textbf{Prior}\\ \textbf{parameters}\end{tabular} \\ \hline
				\multirow{3}{*}{\textbf{I}} & \begin{tabular}[c]{@{}c@{}}Equidistant entry times\\ under equal time trend\end{tabular} & 3 & \begin{tabular}[c]{@{}c@{}}$d_i = d \cdot (i-1)$ $\forall i \ge 1$,\\  $d \in [0, 500]$, \\  with increments of 125\end{tabular} & $\lambda_k = \lambda_0 = 0.15$ $\forall k$ & \begin{tabular}[c]{@{}c@{}}Linear,\\ stepwise,\\ inverted-U\end{tabular} & \begin{tabular}[c]{@{}c@{}}\textbf{Time Machine:} $a_{\tau} = 11.562217$\\ $b_{\tau} = 11.562217$\\ \textbf{MAP Prior:} $1/\sigma^2_{\beta} = 0.001$\\ $1/\sigma^2_{\tau} = 0.002$\end{tabular} \\ \cline{2-7} 
				& Unequal time trends & 3 & \begin{tabular}[c]{@{}c@{}}$d_i = d \cdot (i-1)$ $\forall i \ge 1$,\\   $d = 250$\end{tabular} & \begin{tabular}[c]{@{}c@{}}\textbf{1)} $\lambda_1 \in [-0.05, 0.25]$,\\  with increments of 0.375\\ $\lambda_0 = \lambda_2 = \lambda_3 = 0.1$\\ \textbf{2)} $\lambda_1 = \lambda_2 \in [-0.05, 0.25]$,\\  with increments of 0.375\\ $\lambda_0 = \lambda_3 = 0.1$\end{tabular} & \begin{tabular}[c]{@{}c@{}}Linear,\\ stepwise,\\ inverted-U\end{tabular} & \begin{tabular}[c]{@{}c@{}}\textbf{Time Machine:} $a_{\tau} = 11.562217$\\ $b_{\tau} = 11.562217$\\ \textbf{MAP Prior:} $1/\sigma^2_{\beta} = 0.001$\\ $1/\sigma^2_{\tau} = 0.002$\end{tabular} \\ \cline{2-7} 
				& \begin{tabular}[c]{@{}c@{}}Time Machine and \\ MAP Prior calibration\end{tabular} & 3 & \begin{tabular}[c]{@{}c@{}}$d_i = d \cdot (i-1)$ $\forall i \ge 1$,\\  $d = 250$\end{tabular} & \begin{tabular}[c]{@{}c@{}}$\lambda_k = \lambda_0 \in [-0.15, 0.15]$ $\forall k$,\\  with increments of 0.375\end{tabular} & Stepwise & \begin{tabular}[c]{@{}c@{}}\textbf{Time Machine:} $(a_{\tau}, b_{\tau}): (11.562213, 1156.22134);$ \\ $(4.873556, 121.8389); (11.562217, 11.56222);$ \\ $(1.099121, 0.00010991); (109.912060, 0.00010991)$\\ \textbf{MAP Prior:} $1/\sigma^2_{\beta} \in \{0.001, 1\}$\\ $1/\sigma^2_{\tau} \in \{2, 0.2, 0.002\}$\end{tabular} \\ \hline
				\textbf{II} & \begin{tabular}[c]{@{}c@{}}Non-equidistant \\ entry times under\\ equal time trends\end{tabular} & 4 & \begin{tabular}[c]{@{}c@{}}$d_1 = 0; d_2 = 300; d_4 = 800;$ \\ $d_3 \in [300, 800]$, \\   with increments of 125\end{tabular} & $\lambda_k = \lambda_0 = 0.15$ $\forall k$ & \begin{tabular}[c]{@{}c@{}}Linear,\\ stepwise,\\ inverted-U\end{tabular} & \begin{tabular}[c]{@{}c@{}}\textbf{Time Machine:} $a_{\tau} = 11.562217$\\ $b_{\tau} = 11.562217$\\ \textbf{MAP Prior:} $1/\sigma^2_{\beta} = 0.001$\\ $1/\sigma^2_{\tau} = 0.002$\end{tabular} \\ \hline
				\textbf{III} & \begin{tabular}[c]{@{}c@{}}Unequal time trends centered \\ on the strength of the time \\ trend in the control arm\end{tabular} & 10 & \begin{tabular}[c]{@{}c@{}}$d_i = d \cdot (i-1)$ $\forall i \ge 1$,\\  $d = 300$\end{tabular} & \begin{tabular}[c]{@{}c@{}} $\lambda_0 = \lambda_{10} \in [-0.15, 0.15]$,\\  with increments of 0.375;\\ $\lambda_i \sim N(\lambda_0, 0.5)$ for $0 < i < 10$\end{tabular} & Linear & \begin{tabular}[c]{@{}c@{}}\textbf{Time Machine:} $a_{\tau} = 11.562217$\\ $b_{\tau} = 11.562217$\\ \textbf{MAP Prior:} $1/\sigma^2_{\beta} = 0.001$\\ $1/\sigma^2_{\tau} = 0.002$\end{tabular} \\ \hline
		\end{tabular}}
		\caption{Settings and parameters considered in the simulation study. For a detailed explanation of the settings, see Sect. \ref{sec:simsett}. $K$ refers to the total number of experimental treatment arms, $d$ denotes the difference between entry times of two consecutive treatment arms, and $\lambda$ is the strength of the time trend. Time trend patterns are illustrated in Figure \ref{fig:trend_patterns}; the prior parameters for the Time Machine and MAP approach varied in the simulations are described in Sect. \ref{sec:simmeth} (see equations \eqref{tau}, \eqref{MAP_sigmabeta} and \eqref{MAP_sigmatau}).}
		\label{table_settings}
	\end{table}
\end{landscape}

\subsection{Choice of Priors for the Bayesian Approaches}\label{sec:simmeth}

For the Bayesian approaches, Time Machine model and MAP prior approach, we consider different parameter constellations for the priors to investigate the robustness of the results with respect to design parameter assumptions. 

\begin{itemize}
	\item[] \textbf{Time Machine.} We used bucket sizes of 25 in all settings. For the values of the prior distributions' hyperparameters, we considered precisions of the prior regarding the treatment effect and control response to be 0.001 and 0.001, corresponding to the reciprocal of $\sigma^2_{\eta_0}$ and $\sigma^2_\theta$, and then $\eta_0 \sim N(0,1000)$ and $\theta_k \sim N(0,1000)$.  
	
	We consider different hyperprior distributions for the drift parameter, $\tau\sim Gamma(a_{\tau},b_{\tau})$, depending on the design.     
	In Setting I with equal time trends, we investigate the impact of the choice of value of the prior distribution of the time drift. For this, suppose that at the design stage of the trial, we define the prior of the time drift by assuming a stepwise functional form of the time trend where the strength of the time trend value is a nuisance parameter. We calibrate the values of the prior for $\tau$ by anticipating the expected change between buckets by means of 
	$D_{\text{Expected}}$ and $D_{\text{Maximum}}$ 
	as described in Sect. \ref{sect_tmachine}. The values considered for $D_{\text{Expected}}$ and $D_{\text{Maximum}}$ are  summarised in Table \ref{table_TMpriors}. In all cases, we set  $\iota=0.01$.
	
	In settings II and III, as well as in scenarios with different time trends in Setting I, we set the values of the prior of the time trend corresponding to the assumption of expected change of $D_{\text{Expected}}=1$  and maximal change of $D_{\text{Maximum}}=1.5$. Additional results using $D_{\text{Expected}}=0.01$ and $D_{\text{Maximum}}=0.15$ are to be found in Section B of the Supplementary Material.

	\item[] \textbf{MAP prior approach}. 
	
	Similarly to the Time Machine,   the prior distributions for the treatment effects are $\theta_k \sim N(0,1000)$.  
	The weight used for the non-informative component for the robustification of the MAP prior was set to 0.1 (corresponding to a weight of 0.9 for the MAP component). We used unit-information priors for the weakly-informative mixture component, $p_{\text{non-inf}}$, of the robustified MAP as suggested by \cite{Schmidli2014,Weber2021}.
	
	In Setting I with equal time trends, we consider different cases for the choice of the precision parameter of the half-normal hyperprior for the between-period heterogeneity: $1/\sigma^2_\tau \in \{2, 0.2, 0.002\}$, as well as two different cases for the precision parameter of the normal hyperprior: $1/\sigma^2_\beta \in \{ 0.001, 1 \}$.     
	
	In the remaining cases, the precision $1/\sigma^2_\tau$ was set to 0.002 and the value of the precision parameter of the normal hyperprior to $1/\sigma^2_\beta=0.001$, corresponding to the following prior distribution for the control mean: $\mu_\eta \sim \mathcal{N}(0, 1000)$. Additional results using $1/\sigma^2_\beta = 1$ are to be found in Section B of the Supplementary Material. 
\end{itemize}

\begin{table}
	\centering
	\begin{tabular}{cccc}
		\hline
		\textbf{Expected change} & \textbf{Maximal change} & $a_{\tau}$ & $b_{\tau}$  \\ \hline
		10 & 15 & 11.562213 & 1156.22134 \\ \hline
		5 & 10 & 4.873556 & 121.8389 \\ \hline
		1 & 1.5 & 11.562217 & 11.56222 \\ \hline
		0.01 & 0.15 & 1.099121 & 0.00010991 \\ \hline
		0.001 & 0.015 & 109.912060 & 0.00010991 \\ \hline
	\end{tabular}
	\caption{Considered cases for the calibration of the prior of the time drift for the Time Machine as described in Sect. \ref{sect_tmachine} with $\iota=0.01$. Expected change refers to $D_{\text{Expected}}$ (see \eqref{d_exp}) and Maximal change refers to $D_{\text{Maximum}}$ (see \eqref{d_max}).}
	\label{table_TMpriors}
\end{table} 

\subsection{Results}

\subsubsection{Setting I: Three experimental arms and equidistant entry times}

\begin{figure}[h!]
	\centering
	\includegraphics[width=\textwidth]{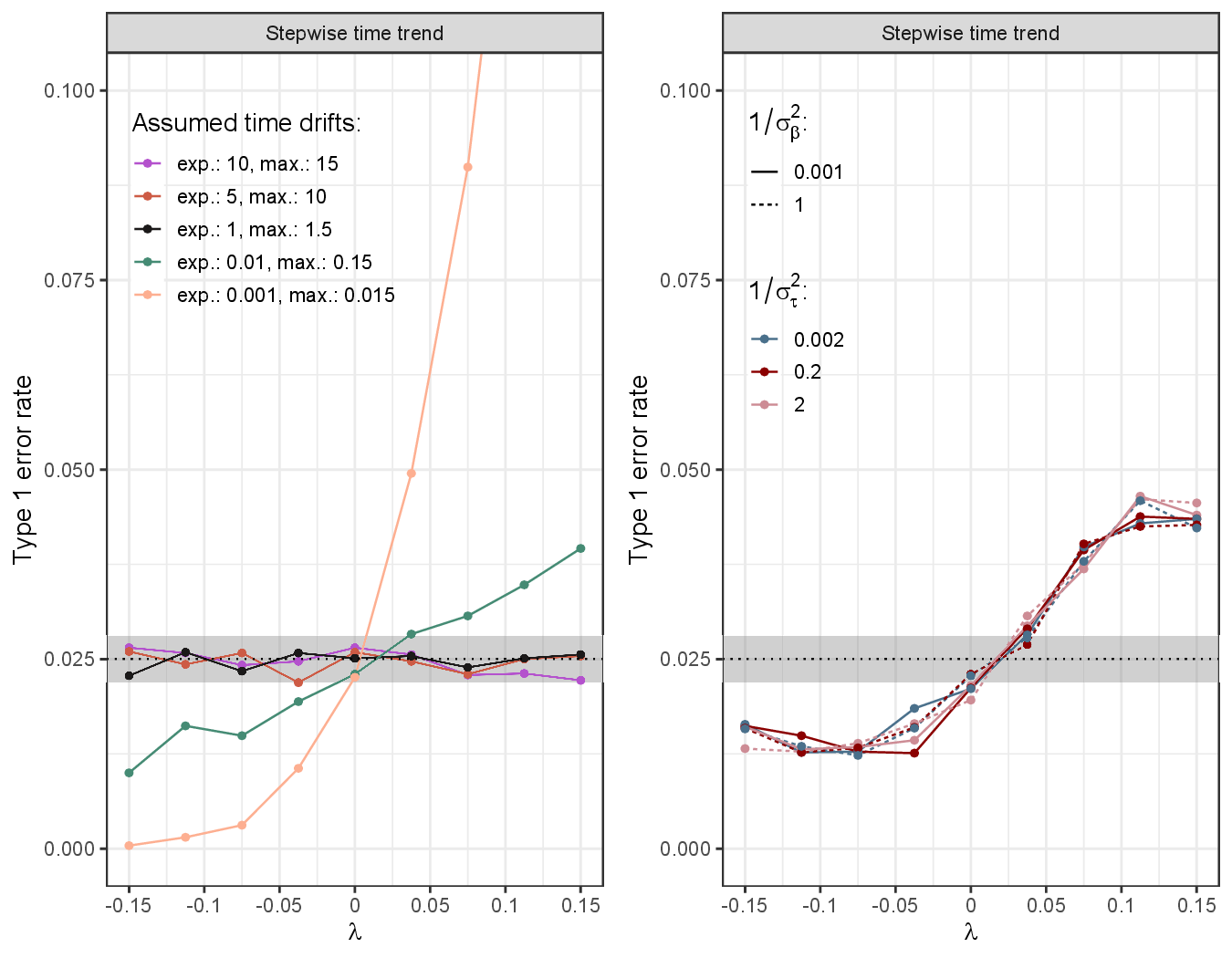}
	\caption{Type 1 error rate of the comparison of arm 3 versus control with respect to the strength of the time trend $\lambda$ in a setting with 3 experimental treatment arms and equal stepwise time trends (with $d=250$) when calibrating the priors for the Time Machine (left panel) and MAP approach (right panel). For the Time Machine, the calibration consists of varying the size of the assumed jump for the Time Machine (see Table \ref{table_TMpriors}); for the MAP, it consists of varying the between-period heterogeneity and precision parameter for the control response. The grey area represents the 95\% prediction interval of the simulated type 1 error rate with 10,000 replications, provided that the true type 1 error rate is 0.025.}
	\label{fig:i_eq_alpha_lambda_TM_main}
\end{figure}

We describe the impact of the calibration of the priors in Figure \ref{fig:i_eq_alpha_lambda_TM_main}. The left plot shows type 1 error curves with respect to the strength of the time trend for different calibrations of the prior of time drift for the Time Machine, and the right plot shows the analogous plot for the calibration of the prior for the MAP approach. For the Time Machine, as expected, for positive time trends, higher type 1 error inflation is found in scenarios in which there was a firmer belief in smooth time trends (i.e. small values of $D_{\text{Expected}}$ -- see, for instance, the curve corresponding to $D_{\text{Expected}}=0.001$). In addition, the type 1 error rate increases with respect to the strength of the time trend in scenarios that used small values for the expected change between time buckets, $D_{\text{Expected}}$. When we assume larger values for the expected change in time drifts in the calibration, then the type 1 error rate is, in general, maintained regardless of the strength of the time trend. When assuming intermediate values as $D_{\text{Expected}}=0.01$, the type 1 error rate is maintained under weak time trends, but for strong trends, there is a type 1 error inflation in case of positive trends, and type 1 error conservatism in case of negative trends, falling below the pre-defined significance level. 
Thus, for the scenarios that follow and also in Settings II and III, we will use the calibration $D_{\text{Expected}}=1$ and $D_{\text{Maximum}}=1.5$ since the resulting prior lead to, in general, type 1 error rate control. In the Supplementary material, we include the results when using $D_{\text{Expected}}=0.01$ and $D_{\text{Maximum}}=0.15$, which results in a mild level of type 1 error inflation under equal time trends across arms, but achieves larger power. 
Similarly, when calibrating the MAP approach, we vary the between-period heterogeneity and variability of responses in the control group, which influence the amount of borrowing. The more heterogeneity and variability assumed, the less borrowing and the more type 1 error control, while more borrowing results in a higher power. As it is well-known \cite{Kopp-Schneider2020}, the risk of type 1 error inflation exists when data is borrowed in such methods, and strict control of type 1 error rate implies that no power gain is possible. Jiao et al. \cite{Jiao2019} also showed that the inflation of the type 1 error rate is bounded in the case of the MAP approach since when concurrent and non-concurrent controls are very dissimilar, the approach does not borrow data from the past and the inflation then decreases. This behaviour can also be observed in Figure \ref{fig:i_eq_alpha_lambda_TM_main} for an increasing $\lambda$.
In the scenarios below, we use an intermediate calibration to serve as a trade-off between error control and power gains.

Next, we investigate the impact of the overlap of experimental treatment arms on the power and type 1 error in trials with equal time trends and equidistant entry times for treatment arms. 

\begin{figure}[h!]
	\centering
	\includegraphics[width=\textwidth]{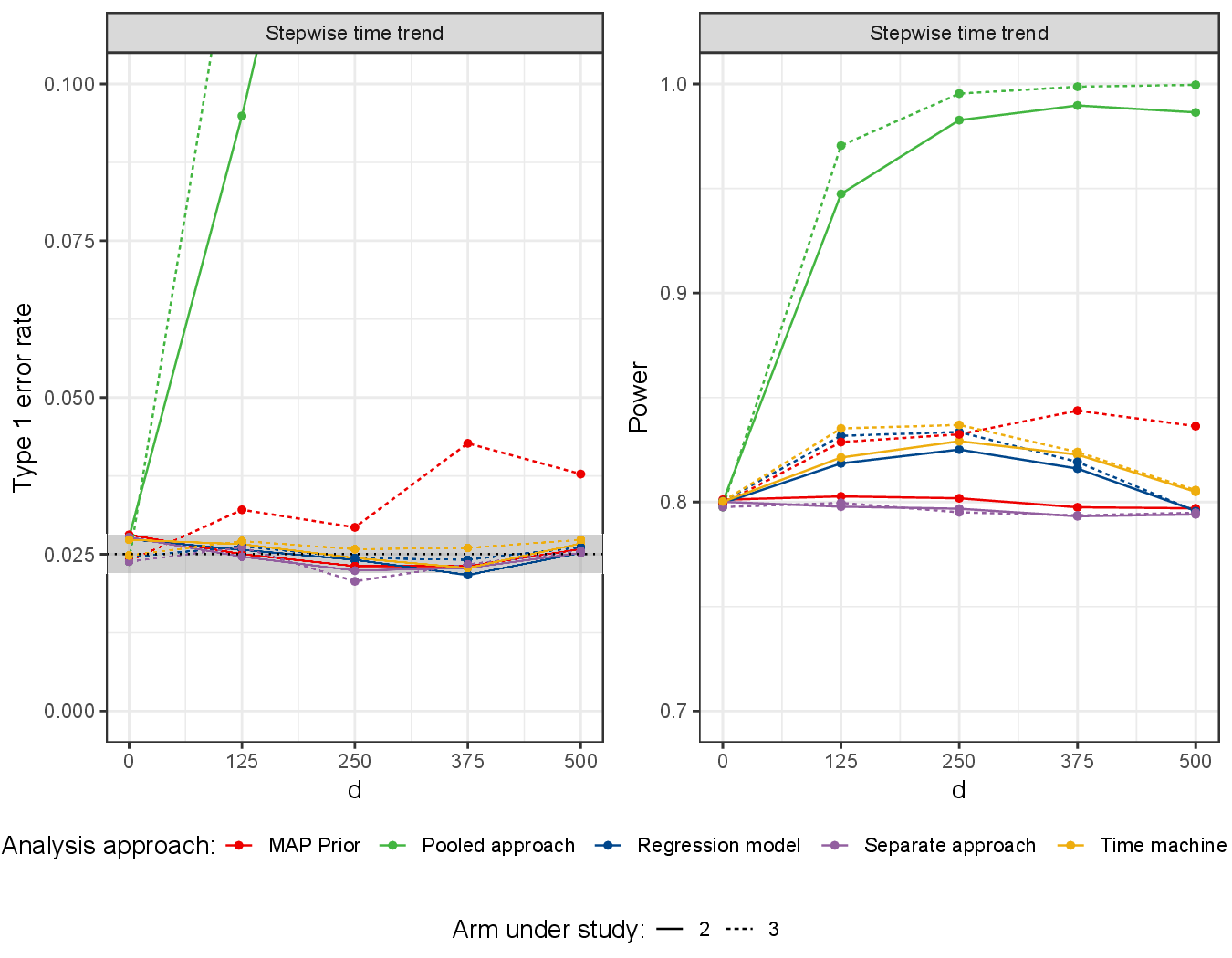}
	\caption{Type 1 error rate (left panel) and power (right panel) with respect to 
		the difference in entry time between consecutive arms $d$ (y-axis) in a setting with three experimental treatment arms (Setting (I), Figure \ref{fig:simdesign}) and equal stepwise time trends ($\lambda_k=0.15$ $\forall k$) according to the analysis approach. The straight and dotted lines refer to the results of arms 2 and 3, respectively. Time Machine calibration consists of $D_{\text{Expected}}=1$ and $D_{\text{Maximum}}=1.5$ according to Table \ref{table_TMpriors}. Prior parameters for the MAP prior are set to $1/\sigma^2_{\beta}=0.001$ and $1/\sigma^2_{\tau}=0.002$  (see Sect. \ref{sec:simmeth}). The grey area represents the 95\% prediction interval of the simulated type 1 error rate with 10,000 replications, provided that the true type 1 error rate is 0.025.}
	\label{fig:i_eq_alpha_d_main}
\end{figure}

The left plot in 
Figure \ref{fig:i_eq_alpha_d_main} shows type 1 error curves with respect to the difference between the entry time of one experimental arm and the next one, $d$. Therefore, the smaller the $d$, the more overlap between experimental treatment arms. Also note that the larger the value of $d$, the larger the size of non-concurrent controls. Thus, the inflation in type 1 error for the pooled analysis, which does not adjust for time drifts, increases with $d$. We can also see a small inflation in the considered scenarios for the MAP approach for increasing $d$. The type 1 error rate is maintained when using the frequentist regression model and the Time Machine (with $D_{\text{Expected}}=1$ and $D_{\text{Maximum}}=1.5$). 
In terms of power, the pooled analysis gives rise to the most powerful method at the cost of the inflation in type 1 error discussed above. 
Time Machine and the frequentist model perform similarly and achieve a power increase as compared to the separate approach. 
When smaller time drifts in the Time Machine prior calibration are assumed (for instance, $D_{\text{Expected}}=0.01$ and $D_{\text{Maximum}}=0.15$), the Time Machine has larger power over the frequentist regression model (see Supplementary material, Figure 3). If assuming larger time drifts, the Time Machine controls the type 1 error but limits borrowing non-concurrent controls. As a result, it leads to similar gains as the frequentist regression model approach, as we can observe in Figure \ref{fig:i_eq_alpha_d_main}.

As for the MAP approach, the power for the Time Machine increases with $d$ in the case of smaller time drift calibration (see Supplementary material, Figure 3). For larger time drift calibration, the Time Machine behaves similarly to the frequentist regression model. In such a case, both models reach their maximum at an intermediate value of $d$ and then decrease (see Figure \ref{fig:i_eq_alpha_d_main}). This is because the Time Machine utilises the non-concurrent data to update the prior of the time drift even if there is no overlap between experimental treatment arms. However, in the frequentist model, only if there is an overlap between arms will there be an estimated time period effect. If not, then the non-concurrent data is not used, and thus, the power coincides with the power of separate trials.

Another point to note is that evidently the later is the entry of the arm under evaluation, the larger the potential inflation of type 1 error (if any) and the larger the increase in power when using non-concurrent controls. We can observe that the type 1 error behaviour is magnified when we compare the results obtained for arm 3 (dotted line) with the results of arm 2  (solid line). This is because treatment arm 3 enters later, and therefore, its corresponding analysis makes more use of non-concurrent data, thus increasing the potential inflation.

\begin{figure}[h!]
	\centering
	\includegraphics[width=\textwidth]{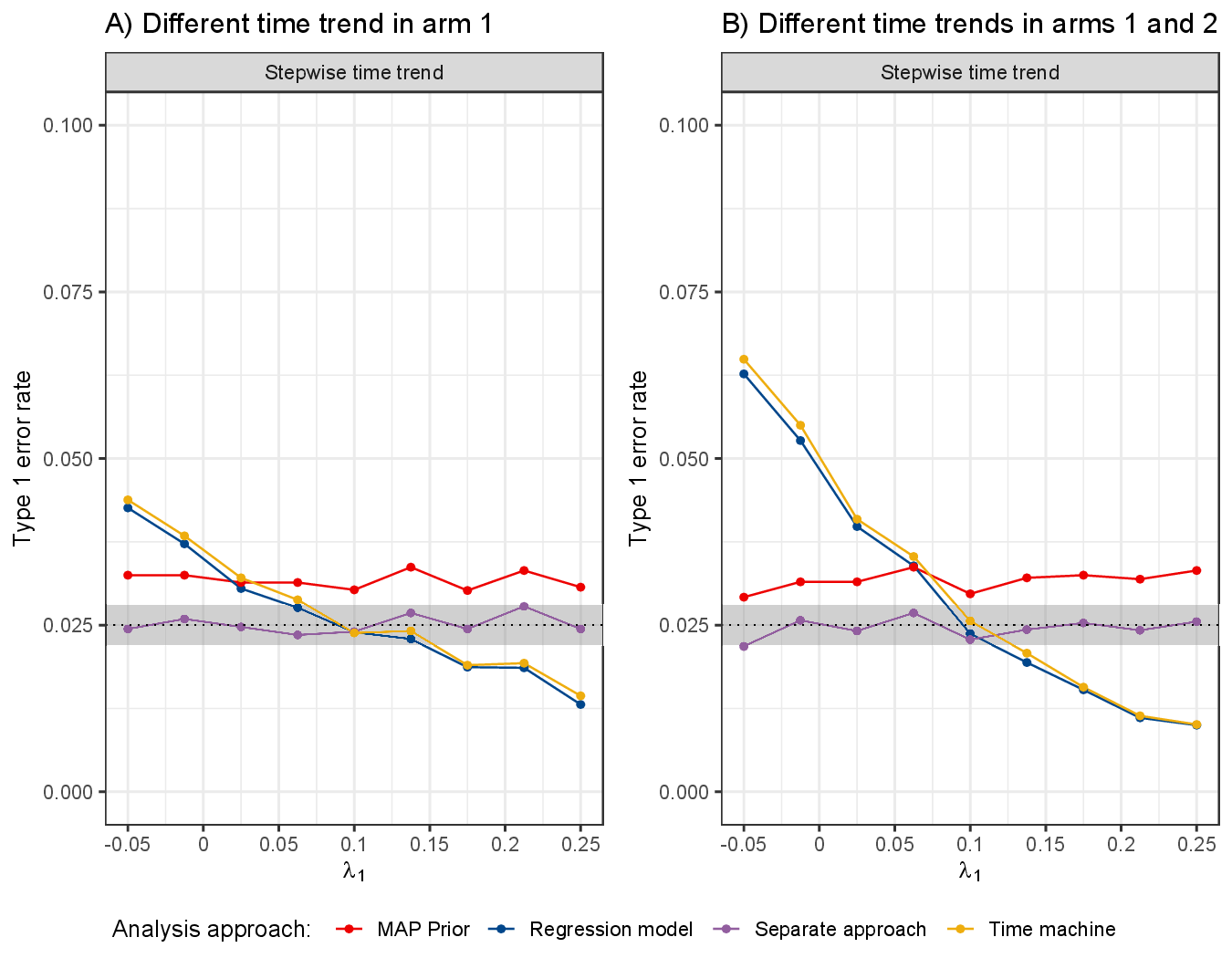}
	\caption{Type 1 error rate of the comparison of arm 3 versus control with respect to the strength of the time trend in arm 1 ($\lambda_1$) in  Setting (I) with 3 experimental treatment arms under different stepwise time trends according to the analysis approach. On the left panel,   (A)  only the time trend for arm 1 differs from the rest ($\lambda_0 = \lambda_2 = \lambda_3 = 0.1$); on the right panel, (B) time trends for arms 1 and 2 ($\lambda_1 = \lambda_2 $)  differ from time trends in control and arm 3  ($\lambda_0 = \lambda_3 = 0.1$). Time Machine calibration consists of $D_{\text{Expected}}=1$ and $D_{\text{Maximum}}=1.5$ according to Table \ref{table_TMpriors}. Prior parameters for the MAP prior are set to $1/\sigma^2_{\beta}=0.001$ and $1/\sigma^2_{\tau}=0.002$ (see Sect. \ref{sec:simmeth}). The grey area represents the 95\% prediction interval of the simulated type 1 error rate with 10,000 replications, provided that the true type 1 error rate is 0.025.}
	\label{fig:i_diff1_alpha_lambda_main}
\end{figure}

As mentioned above, the frequentist model (Sect. \ref{sect_freqmods}) and the Time Machine (Sect. \ref{sect_tmachine}) assume that the effect of time is additive and affects all arms in the trial equally. Here, we also investigate the robustness of the methods when the assumption of equal time trends is violated. Figure \ref{fig:i_diff1_alpha_lambda_main} shows the type 1 error rate results in trials with different time trends.   The left panel refers to a trial where only time trends in arm 1 are different to the others, while the right panel refers to when the time trends are different in arms 1 and 2. We can observe how the frequentist regression and Time Machine fail to control the type 1 error rate. Inflation becomes more pronounced the more different the trend in arm 1 is compared to the rest (i.e., as larger $\lambda_0-\lambda_1$) (see left plot). We can also see that the pattern worsens when both arm 1 and arm 2 are different (see right plot). Figure \ref{fig:i_diff1_alpha_lambda_main} presents the case under stepwise time trends. However, it is worth noting that such inflation is lower when the trends are linear or even inverted-U time trends (see Supplementary Material, Figure 2). 

\subsubsection{Setting II: Four experimental arms and non-equidistant entry times}

\begin{figure}[h!]
	\centering
	\includegraphics[width=0.75\textwidth]{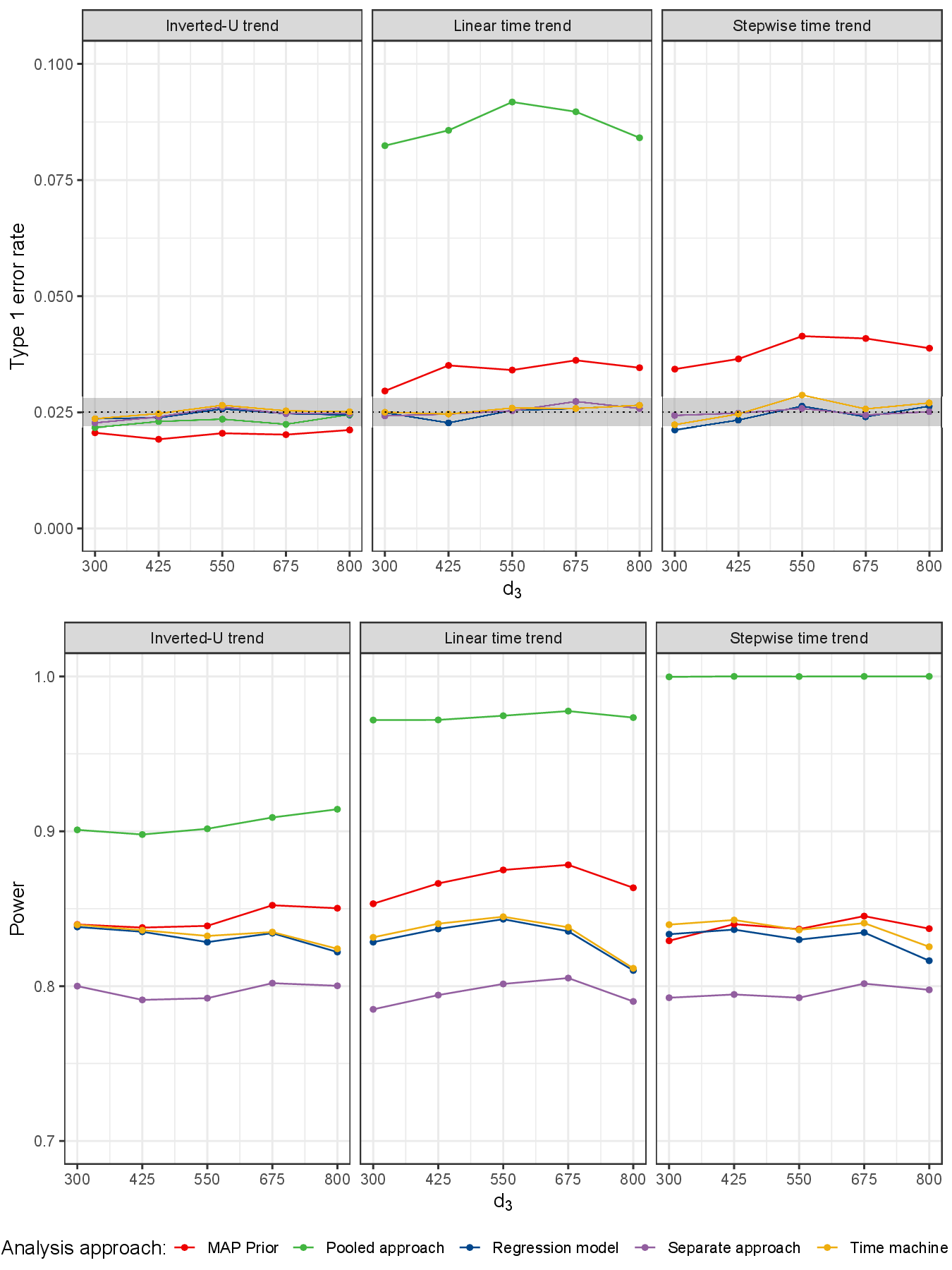}
	\caption{Type 1 error rate (first row) and power (second row) of the comparison of arm 4 against control for different time trend patterns (equal time trends of $\lambda_k=0.15$, $\forall k$) and with respect to the timing of adding the treatment arm 3 ($d_3$) and according to the analysis approach. Results correspond to setting II with four experimental treatment arms, where arms 2 and 4 enter after $d_2=300$ and $d_4=800$ participants have been allocated to the trial, respectively. Time Machine calibration consists of $D_{\text{Expected}}=1$ and $D_{\text{Maximum}}=1.5$ according to Table \ref{table_TMpriors}. Prior parameters for the MAP prior are set to $1/\sigma^2_{\beta}=0.001$ and $1/\sigma^2_{\tau}=0.002$  (see Sect. \ref{sec:simmeth}). The grey area represents the 95\% prediction interval of the simulated type 1 error rate with 10,000 replications, provided that the true type 1 error rate is 0.025. Note that results using the pooled approach are not visible in the top right figure, as they go outside the scale set on the y-axis.}
	\label{fig:ii_eq_alpha_d_main}
\end{figure}

Next, we evaluate trials with non-equidistant arm entry under equal time trends across arms. In particular, we evaluate a four-arm trial where arm 3 enters at some point between arms 2 and 4. The first row in Figure \ref{fig:ii_eq_alpha_d_main} shows the type 1 error with respect to the entry time of arm 3, $d_3$, according to the pattern of the time trend. We can see that Bayesian methods are affected by such a change in this case, and the type 1 error rate is considerably affected, especially in stepwise time trends. 
In the case of the Time Machine, this loss of control is especially evident when assuming medium or small time drifts in the prior calibration (see, for instance, the results when using $D_{\text{Expected}}=0.01$ and $D_{\text{Maximum}}=0.15$ in the Supplementary Material, Figure 5). This is because, on the one hand, the effect of the time trend is greater with such a change since there are more jumps in a shorter time. Also, in the case of $d_2=d_3$, even the jump is doubled, which means that there is a greater difference between the concurrent and non-concurrent controls than in the previously discussed settings. On the other hand, the Time Machine assumes that the time drift will be similar over time, which is not the case here, as the time drifts depend on the entry times of the arms in the stepwise time trends, and arms do not enter in an equidistant manner. For the MAP approach, we can see that the inflation remains constant with respect to the entry time $d_3$, but increases under stepwise time trends. The reason is that when dealing with linear or inverted-U trends, the time trend increases gradually, whereas the changes are more abrupt and severe with stepwise, resulting in more inflation.  

When inspecting the power in Figure \ref{fig:ii_eq_alpha_d_main} (second row), we can see that the power obtained when using the Time Machine is slightly higher than the power using the frequentist regression, in which case the type 1 error rate was controlled when using the Time Machine but consistently larger than the one obtained by using the frequentist model. 

\subsubsection{Setting III: Ten experimental arms and equidistant entry times}

\begin{figure}[h!]
	\centering
	\includegraphics[width=\textwidth]{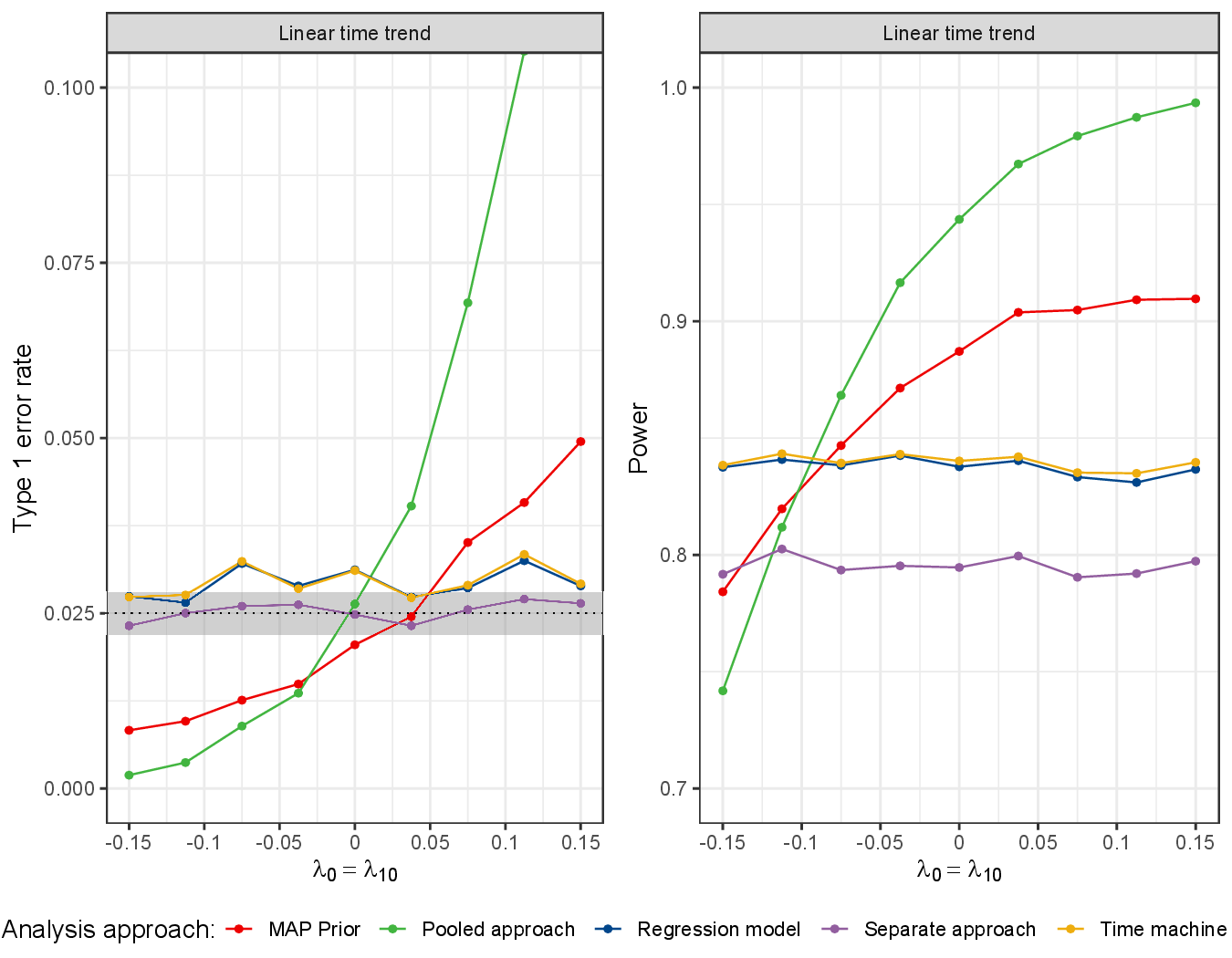}
	\caption{Type 1 error rate (left panel) and power (right panel) of the comparison of arm 10 against control with respect to the analysis approach in a setting III with ten experimental treatment arms and different linear time trends, where the time trend for the first 9 arms is sampled from a normal distribution $N(\lambda_0, 0.5)$, and the control and arm 10 have equal time trends ($\lambda_0 = \lambda_{10}$). In this setting, a treatment arm $k$ enters after $300 \cdot (k-1)$ participants have been allocated to the trial. Time Machine calibration consists of $D_{\text{Expected}}=1$ and $D_{\text{Maximum}}=1.5$ according to Table \ref{table_TMpriors}. Prior parameters for the MAP prior are set to $1/\sigma^2_{\beta}=0.001$ and $1/\sigma^2_{\tau}=0.002$  (see Sect. \ref{sec:simmeth}). The grey area represents the 95\% prediction interval of the simulated type 1 error rate with 10,000 replications, provided that the true type 1 error rate is 0.025.}
	\label{fig:iii_ii_rand_alpha_lambda_main}
\end{figure}

Here, we assume a longer trial, with ten experimental treatment arms. We assume that the strength of the time trends may vary between treatments but that they are equally distributed according to a normal distribution with a mean equal to the time trend strength in the control group and equal variance. Under this setting, we can see in Figure \ref{fig:iii_ii_rand_alpha_lambda_main} that the type 1 error rate for arm 10 vs control is maintained when using frequentist regression, and so its performance is similar to that of the separate approach. This is because the differences in time trends across arms are averaged out, and the effect of different time trends gets diluted. This is also the case for the Time Machine when using $D_{\text{Expected}}=1$, but not with $D_{\text{Expected}}=0.01$ (see Supplementary material, Figure 6). 
For the later case of the Time Machine and also for the MAP, there is a perceived loss of type 1 error rate control, leading to either conservative tests or inflation of the type 1 error rate depending on whether the intensity in the control group is negative or positive, respectively.


\section{Discussion}\label{sect_discuss}


In platform trials, arms enter and leave while the control arm continues enrolling participants. Adding a new arm in a platform trial comes with pre-existing data from control participants, referred to as non-concurrent controls. The development of methods to utilise non-concurrent controls in platform trials is an area undergoing rapid evolution, engaging both academia and regulatory bodies. Deciding whether to incorporate these controls into the analysis of treatment versus control involves weighing advantages and disadvantages. Utilising non-concurrent controls offers the potential benefit of providing information about the control arm, leading to more precise estimates and enhanced power for comparing treatment effectiveness. However, it is essential to recognise that non-concurrent controls do not come from the same randomised process as the concurrent controls but rather stem from an earlier time of the trial. This temporal distinction between the earlier and current parts of the trial can introduce confounding factors into the analysis, especially if time trends are present. Current research is principally focused on modeling time effects, exploring diverse models, identifying necessary assumptions for robust performance, and investigating the worst-case scenarios in case those assumptions are not met. In this work, we primarily aimed to compare current methods to incorporate non-concurrent controls. But we also extend existing methods for treatment-control comparisons incorporating non-concurrent control data. The Time Machine approach \cite{Saville2022} was originally proposed for trials with binary endpoints, and here we extended it to continuous endpoints. In addition, we have extended the Bayesian meta-analytic-predictive (MAP) approach for utilizing historical controls, to use it for non-concurrent controls in platform trials. 

In this work, we investigate frequentist and Bayesian modelling approaches that model the time trend and Bayesian strategies that limit the borrowing level depending on the heterogeneity between concurrent and non-concurrent controls. In particular, we consider the frequentist model-based adjustments that were proposed in \cite{BofillKrotka2022}, the MAP approach, and the Bayesian Time Machine. 
When evaluating the statistical power and the  type 1 error rate for individual treatment-control comparisons, we examine the impact of time trends in each method under different patterns for the time trends and the role of the overlap between arms. 
When considering time trends, two important points need to be taken into account. The first would be the interaction of the trends with the treatment arms, i.e., whether we consider that the time trends will affect all arms similarly or not. Second, model-based approaches assume that the time trend is on the same scale as the response and that it is additive with respect to the model. This point might be especially relevant in trials with binary or survival endpoints, and it is worth further study in this direction. 

For trials with potentially different time trends across arms, there have been recent proposals to ease the assumption of equal time trends by taking into account the interaction between treatment and time as a random factor. This approach helps to lower the inflation of type 1 error in scenarios where there are different time trends between treatment groups. However, it is still not possible to maintain strict error control \cite{krotka2024}. In case of interactions of the treatment effect with time, already the interpretation in a simple two-arm trial comparing a single experimental treatment to control can become tricky. E.g., assume an improvement in the control arm over time but none in the experimental arm. If there is enough difference early in the trial, the final results might yield a statistically significant difference. But actually, at the end of the trial, there might be no advantage over the control arm left. In standard trials, this time aspect is usually not investigated and, therefore, not discussed. But, platform trials now offer the opportunity to reveal such issues, especially when considering the utilization of non-concurrent control data for decision-making. 
We have seen that interactions of the treatment effect with time affect modelling-based methods, like frequentist and Time Machine, which may lose control of type 1 error. Also, we saw that the effect of unequal time trends is more pronounced when more than one arm presents unequal time trends, and in addition, those arms do have equal time trends between them. On the other hand, if the trends follow the same pattern and the strength of the trend is different in the different arms but the same on average, the frequentist regression keeps the type 
1 error rate under control.

Regarding overlap between experimental treatment arms, intermediate overlaps result in the highest gains in power for the frequentist model. This is also true for the Time Machine model when considering priors controlling the type 1 error,  where the results are similar to those of the frequentist model. If there is no overlap, the frequentist model loses power and the Time Machine can lose type 1 error control. The MAP approach does not directly adjust for time trends and, in general, as was the case in trials incorporating external controls, does not control the type 1 error in the presence of time trends \cite{Kopp-Schneider2020}. Both MAP approach and Time Machine rely on the assumptions regarding the prior distributions.  We have seen that in the case of Time Machine, if one restricts to priors that give rise to type 1 error rate control, the gain in power is minimal compared to the frequentist regression model. If one chooses priors potentially leading to more inflation of type 1 error rate, one can gain more power than when using the frequentist regression.
There is extensive literature on Bayesian methods for using historical controls (see \cite{Viele2014} for a review of methods); we have chosen MAP as it is one of the most well-known and widely used. Other methods, such as power prior approaches, could be considered and further explored in this context.
When considering the MAP prior approach, we keep the original idea of this approach and consider each period as if it were a different source of information. This way of constructing the MAP does not, however, take into account the order of the periods. One could consider extending the approach by including the time variable or stochastic order of the periods analogously to the Time Machine.

Modeling approaches, such as the frequentist regression and Bayesian Time Machine, align with the recent Food and Drug Administration (FDA) draft guidance on Master Protocols \citep{FDA2023masterprotocol}, which suggests the use of stratified analyses to avoid bias caused by time trends. If and how non-concurrent control data will be used in platform trials already needs a pre-specification in the master protocol \citep{gidh2024developing, nguyen2024reg} without any knowledge of the control arm when adding later treatment arms. The justification should ideally also include a discussion on which assumptions have to be taken for the analysis considered. Regardless of the chosen method, a clear understanding of the required assumptions is crucial. If these assumptions are met, results can benefit from improvement in the precision of estimates and gain in statistical power. Having a thorough awareness of risks is imperative before taking any. While taking risks may be acceptable in some cases, it is important to quantify them beforehand. Moreover, these assumptions play a pivotal role in regulatory interactions. Proposing a design that utilises non-concurrent controls requires a precise statement of the assumptions being made, an accurate assessment of the robustness of the proposed method with respect to the scenarios to be envisaged in the specific indication and an evaluation of the risks to be taken and potential gains.  

\section*{Supplementary Material}

Additional results from the simulation study. The GitHub repository 
(\url{https://github.com/pavlakrotka/NCC_MethodsComp}) 
contains the R code to reproduce the results of the simulation study.

\section*{Acknowledgments}

EU-PEARL (EU Patient-cEntric clinicAl tRial pLatforms) project has received funding from the Innovative Medicines Initiative (IMI) 2 Joint Undertaking (JU) under grant agreement No 853966. This Joint Undertaking receives support from the European Union’s Horizon 2020 research and innovation programme and EFPIA and Children’s Tumor Foundation, Global Alliance for TB Drug Development non-profit organisation, Spring works Therapeutics Inc. This publication reflects the authors’ views. Neither IMI nor the European Union, EFPIA, or any Associated Partners are responsible for any use that may be made of the information contained herein.

\noindent The Paul Ehrlich Institute receives funding exclusively from the EU Commission.

\noindent This research was funded in whole, or in part, by the Austrian Science Fund (FWF) [ESP 442 ESPRIT-Programm]. For the purpose of open access, the author has applied a CC BY public copyright licence to any Author Accepted Manuscript version arising from this submission.

\section*{Conflict of Interest}

Dominic Magirr declares a competing interest as an employee of Novartis Pharma AG. Peter Jacko and Tom Parke  are employees of Berry Consultants, a consulting company that specialises in the design, conduct, and analysis of Bayesian and adaptive clinical trials. The rest of the authors declare that they have no competing interests regarding the content of this article.


\bibliography{refs}

\newpage

\newpage

\includepdf[pages=-]{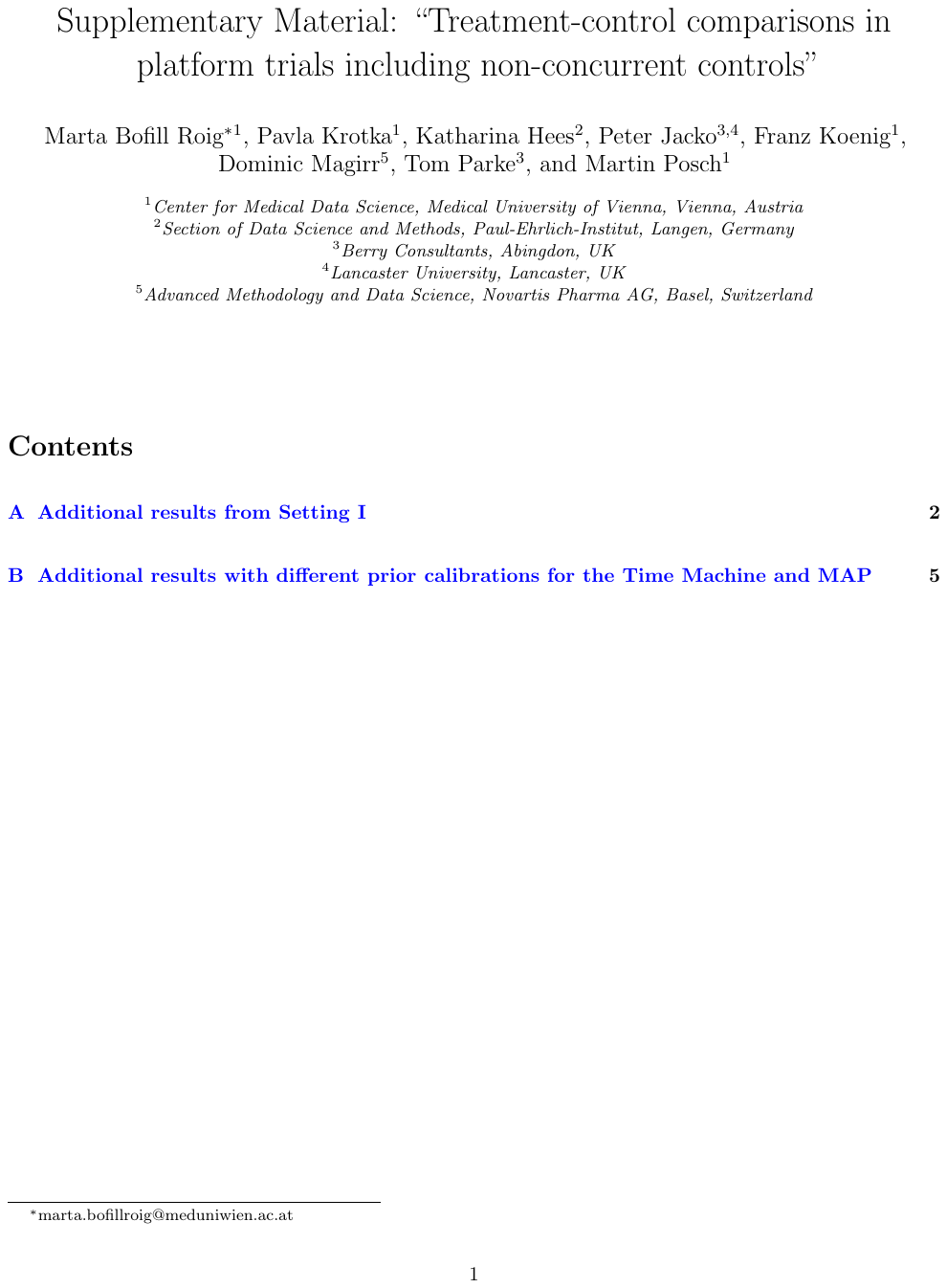}

\clearpage

\end{document}